\documentclass[twocolumn,showpacs,preprintnumbers,amsmath,amssymb,groupedaddress]{revtex4}

\usepackage{graphicx}
\usepackage{graphicx}
\usepackage{here}
\begin{document}
\title{Valence changes associated with the metal-insulator transition 
in Bi$_{1-x}$La$_x$NiO$_3$}

\author{H. Wadati}
\affiliation{Department of Physics and Department of Complexity 
Science and Engineering, University of Tokyo, 
Kashiwanoha 5-1-5, Kashiwa, Chiba 277-8561, Japan}

\author{M. Takizawa}
\affiliation{Department of Physics and Department of Complexity 
Science and Engineering, University of Tokyo, 
Kashiwanoha 5-1-5, Kashiwa, Chiba 277-8561, Japan}

\author{T. T. Tran}
\affiliation{Department of Physics and Department of Complexity 
Science and Engineering, University of Tokyo, 
Kashiwanoha 5-1-5, Kashiwa, Chiba 277-8561, Japan}

\author{K. Tanaka}
\affiliation{Department of Physics and Department of Complexity 
Science and Engineering, University of Tokyo, 
Kashiwanoha 5-1-5, Kashiwa, Chiba 277-8561, Japan}

\author{T. Mizokawa}
\affiliation{Department of Physics and Department of Complexity 
Science and Engineering, University of Tokyo, 
Kashiwanoha 5-1-5, Kashiwa, Chiba 277-8561, Japan}

\author{A. Fujimori}
\affiliation{Department of Physics and Department of Complexity 
Science and Engineering, University of Tokyo, 
Kashiwanoha 5-1-5, Kashiwa, Chiba 277-8561, Japan}

\author{A. Chikamatsu}
\affiliation{Department of Applied Chemistry, University of Tokyo, 
Bunkyo-ku, Tokyo 113-8656, Japan}

\author{H. Kumigashira}
\affiliation{Department of Applied Chemistry, University of Tokyo, 
Bunkyo-ku, Tokyo 113-8656, Japan}

\author{M. Oshima}
\affiliation{Department of Applied Chemistry, University of Tokyo, 
Bunkyo-ku, Tokyo 113-8656, Japan}

\author{S. Ishiwata}
\affiliation{Institute for Chemical Research, Kyoto University, 
Uji, Kyoto 611-0011, Japan}

\author{M. Azuma}
\affiliation{Institute for Chemical Research, Kyoto University, 
Uji, Kyoto 611-0011, Japan}

\author{M. Takano}
\affiliation{Institute for Chemical Research, Kyoto University, 
Uji, Kyoto 611-0011, Japan}

\date{\today}
\begin{abstract}
Perovskite-type BiNiO$_3$ is an insulating 
antiferromagnet in which a charge disproportionation occurs 
at the Bi site. 
La substitution for Bi suppresses the charge disproportionation 
and makes the system metallic. 
We have measured the photoemission and x-ray 
absorption (XAS) spectra of Bi$_{1-x}$La$_{x}$NiO$_{3}$ to investigate 
how the electronic structure changes with La doping. 
From Ni $2p$ XAS, we observed an increase of the valence of
Ni from 2+ toward 3+. 
Combined with the core-level photoemission study, 
it was found that 
the average valence of Bi remains $\sim 4+$ and 
that the Ni valence behaves as $\sim (2+x)+$, that is, 
La substitution results in hole doping at the Ni sites. 
In the valence-band photoemission spectra, 
we observed a Fermi cutoff for $x>0$, 
consistent with the metallic behavior of the La-doped compounds. 
The Ni $2p$ XAS, Ni $2p$ core-level photoemission, 
and valence-band photoemission spectra 
were analyzed by configuration-interaction 
cluster-model calculation, 
and the spectral line shapes 
were found to be consistent with the gradual 
Ni$^{2+} \rightarrow$ Ni$^{3+}$ valence change. 
\end{abstract}
\pacs{71.28.+d, 71.30.+h, 79.60.-i, 78.70.Dm}
\maketitle
\section{Introduction}
Perovskite-type $3d$ transition-metal oxides exhibit various interesting 
physical properties, 
such as metal-insulator (MI) transition, 
colossal magnetoresistance (CMR), 
and ordering of spin, charge, and orbitals \cite{rev}. 
Among them, the charge-transfer-type $R$NiO$_3$ ($R=$ rare earth) 
forms a prototypical system in which 
an MI transition occurs in a systematic 
manner, namely, as a function of the radius of the $R$ ion 
and hence of the one-electron 
bandwidth \cite{Torrance,Nirev}. The least distorted LaNiO$_3$ is a paramagnetic
metal, whereas more distorted $R$NiO$_3$ with a smaller $R$ ion 
becomes an antiferromagnetic insulator. Some $R$NiO$_3$ with 
an $R$ ion of intermediate size shows a metal-insulator transition 
as a function of temperature, too. 

In this context, 
BiNiO$_{3}$ had been expected to be metallic, if synthesized, 
because the ionic radius of
Bi$^{3+}$ is larger than that of La$^{3+}$. 
Recently, Ishiwata {\it et al.} succeeded in synthesizing 
BiNiO$_3$ under a high pressure of 
6 GPa \cite{Ishiwata1}. 
Contrary to the above expectation, BiNiO$_{3}$ was found to be 
an insulating 
antiferromagnet with localized spins of $S=1$. 
X-ray powder diffraction (XRD) study revealed 
that the Bi ions were charge-disproportionated 
into ``Bi$^{3+}$'' and ``Bi$^{5+}$''. 
Thus the oxidation state of the Ni ion was concluded to be 
$2+$ rather than $3+$. 
To the authors' knowledge, 
such a charge disproportionation at the $A$-site in the perovskite structure 
has never been reported so far. 
Since Bi has a deep 6$s$ level, and such a high valence state 
as ``Bi$^{5+}$''at A-site would not be stable from the viewpoint of the
Madelung potential \cite{Ma}, 
the realistic charge configuration of Bi$^{5+}$ may 
be expressed as Bi$^{3+}$$\underline{L}^2$, where $\underline{L}$ 
denotes a hole in the O $2p$ band. 
Subsequently, it was reported that the substitution of La for Bi suppressed 
the charge 
disproportionation and made the system conducting \cite{BLNO}. 
Therefore, an interesting question for this system is 
how the electronic structure changes from the charge-disproportionated 
insulating BiNiO$_3$ to the metallic LaNiO$_3$ with La substitution. 
One possible senario is that the La substitution suppresses 
the charge disproportionation and changes the average valence of Bi 
from $4+$ to $3+$. If so, the valence of Ni would become $3+$, 
and the Ni atoms become responsible for the metallic conductivity. 
Another senario is that the La substituion suppresses 
the charge disproportionation but the average valence of Bi remains 
$4+$. If so, the valence of Ni becomes $(2+x)+$, and both the Bi 
and Ni sites contribute to the metallic conductivity. 
To settle this question, 
we have investigated the electronic structure of
Bi$_{1-x}$La$_{x}$NiO$_{3}$ by photoemission and x-ray 
absorption spectroscopy (XAS). 
Our result seems to support the latter scenario.
\section{Experiment}
\begin{figure}
\begin{center}
\includegraphics[width=6cm]{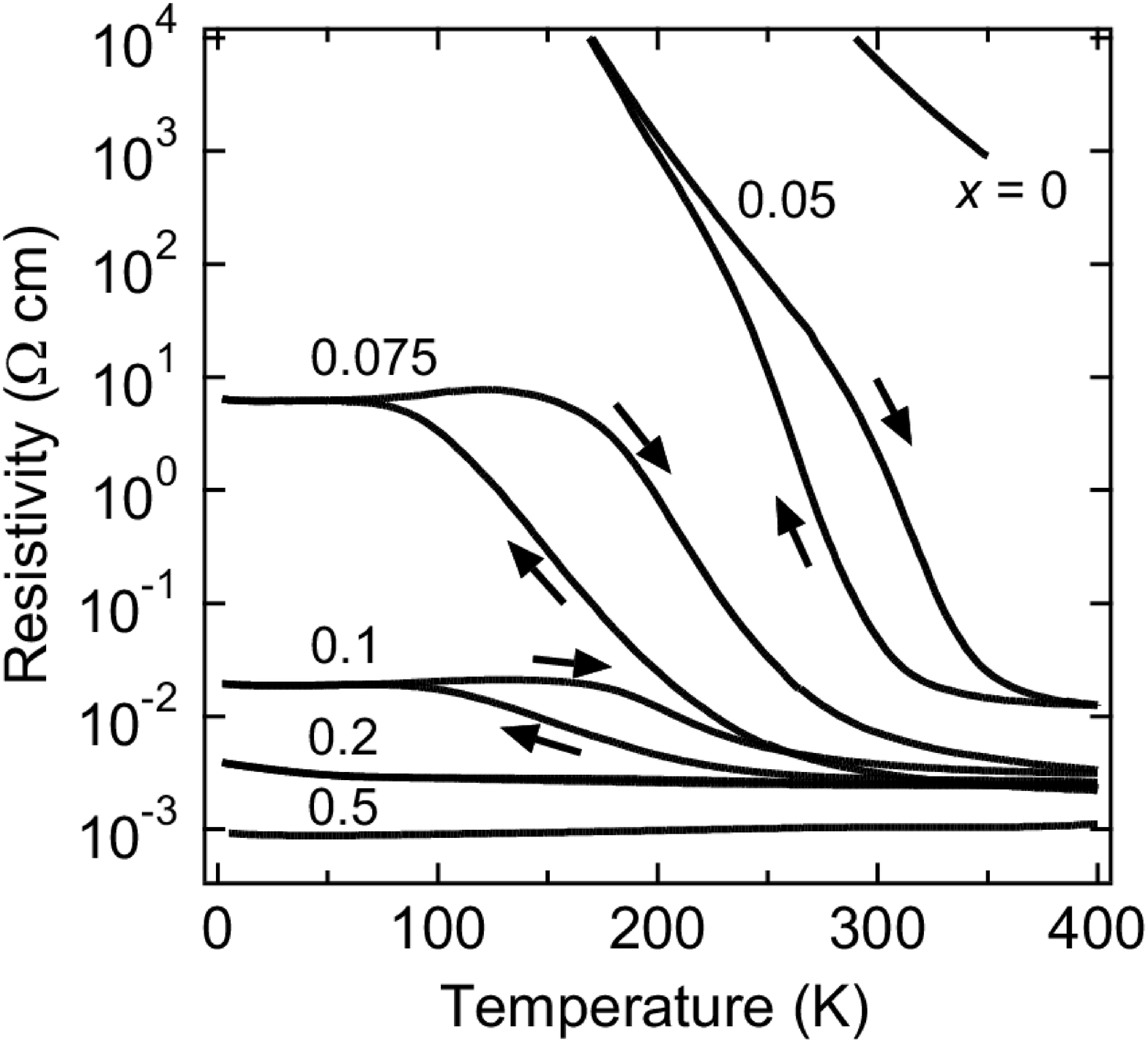}
\caption{Electrical resistivity of Bi$_{1-x}$La$_x$NiO$_3$ \cite{BLNO}.} 
\label{pt}
\end{center}
\end{figure}

Preparation and characterization of polycrystalline 
Bi$_{1-x}$La$_{x}$NiO$_{3}$ ($x=0$, 0.05, 0.1, 0.2) 
are described elsewhere \cite{Ishiwata1,BLNO}. 
Figure \ref{pt} shows the electrical resistivity of 
Bi$_{1-x}$La$_x$NiO$_3$ \cite{BLNO}. 
For $x=0$ (BiNiO$_3$), the system is insulating. 
Electrical resistivity decreases with La substitution. 
For $x=0.05$, 0,075, and 0.1, a broad MI transition occurs 
as a function of temperature. For $x=0.2$ and 0.5, 
the system is metallic. 
All the experiments except for 
ultraviolet photoemission spectroscopy (UPS) 
were performed at BL-2C of Photon Factory (PF), 
High Energy Accelerators Research Organization (KEK). 
UPS measurements were performed 
using a high-flux discharging lamp with a troidal 
grating monochromator. 
The photoemission and x-ray absorption measurements were performed 
under an ultrahigh vacuum 
of $\sim 10^{-10}$ Torr at room temperature. 
The photoemission spectra were measured using 
a Scienta SES-100 electron-energy analyzer. 
At KEK-PF, the total energy resolution was about 200-500 meV depending on the photon energy, whereas for UPS measurments 
it was set to about 10 meV and 50 meV 
for the He I (21.2 eV) and He II (40.8 eV) 
light sources, respectively. 
The Fermi level ($E_F$) position was determined by measuring gold 
spectra. 
The XAS spectra were measured by the total-electron-yield method. 
Clean surfaces of the samples were obtained by repeated {\it in-situ} 
scraping with a diamond file.

\section{Results and discussion}
\subsection{Ni $2p$ x-ray absorption and Ni $2p$ core-level photoemission spectra}
\begin{figure}
\begin{center}
\includegraphics[width=9cm]{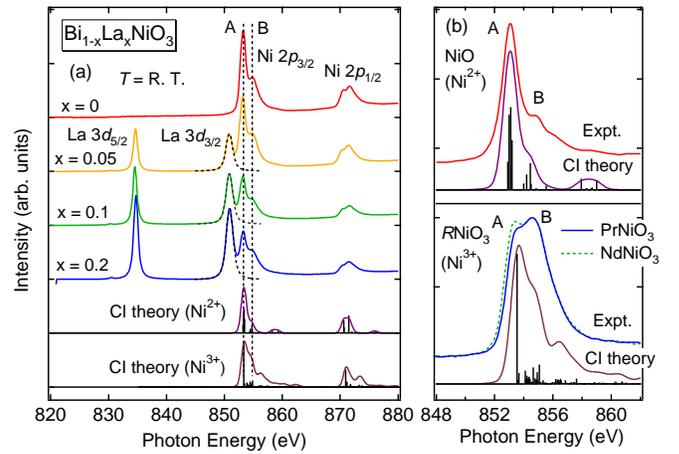}
\caption{(Color online) Ni $2p$ XAS 
spectra and 
their CI cluster-model analyses. 
(a) Ni $2p$ XAS spectra of Bi$_{1-x}$La$_x$NiO$_3$. 
The calculated spectra 
assuming Ni$^{2+}$ and Ni$^{3+}$ are 
presented at the bottom. The 
dashed lines represent the La $3d_{3/2}$ absorption 
peaks. 
(b) Ni $2p$ XAS spectra of NiO and $R$NiO$_3$ from Ref 
\cite{Medarde}.} 
\label{Ni}
\end{center}
\end{figure}

\begin{figure}
\begin{center}
\includegraphics[width=7.5cm]{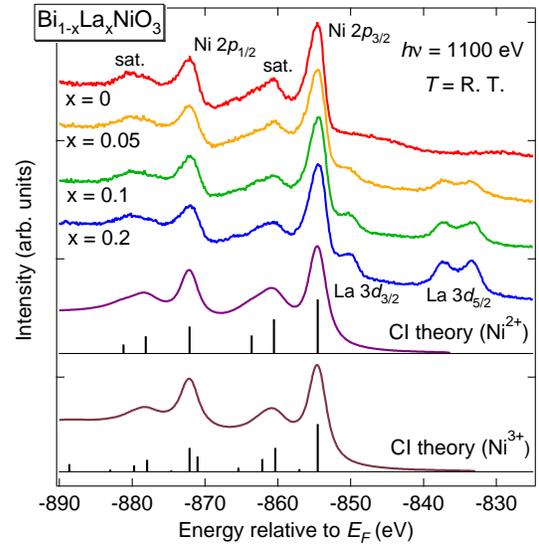}
\caption{(Color online) Ni $2p$ core-level photoemission spectra of 
Bi$_{1-x}$La$_x$NiO$_3$ and 
their CI cluster-model analyses. 
The calculated spectra 
assuming Ni$^{2+}$ and Ni$^{3+}$ are presented
at the bottom.} 
\label{NiXPS}
\end{center}
\end{figure}

Figure \ref{Ni} (a) shows the Ni $2p$ XAS spectra of Bi$_{1-x}$La$_x$NiO$_3$. 
The Ni $2p_{3/2}$ peak 
overlaps the La $3d_{3/2}\rightarrow 4f$ 
absorption peak. 
As reported previously 
\cite{Medarde}, 
the Ni $2p$ XAS spectra of $R$NiO$_3$ (Ni$^{3+}$) are very different 
from that of NiO (Ni$^{2+}$): 
The $2p_{3/2}$ spectrum consists of two peaks labelled as A and B, as
shown in Fig.~\ref{Ni} (a). 
The lower energy peak A is stronger than the higher energy one B 
for NiO, whereas the two peaks have similar intensities in 
$R$NiO$_3$ (as shown in Fig.~\ref{Ni}(b)). 
The spectrum of $x=0$ is 
similar to that of Ni$^{2+}$, which is consistent with the report that 
the valence of Ni is 2+ in BiNiO$_3$ \cite{Ishiwata1}. 
With increasing $x$, 
the relative intensity of the higher energy peak increases, 
indicating that the valence of Ni gradually increases from Ni$^{2+}$ 
toward Ni$^{3+}$. 
Figure \ref{NiXPS} shows the Ni $2p$ photoemission spectra of Bi$_{1-x}$La$_x$NiO$_3$. 
As in the case of XAS, the Ni $2p_{3/2}$ core level overlaps the 
La $3d_{3/2}$ core level. 
The Ni $2p$ photoemission spectra do not change with $x$, 
consistent with the reports that 
the line shapes of the Ni $2p$ photoemission spectra of $R$NiO$_3$ 
(Ni$^{3+}$ or Ni$^{2+}$$\underline{L}$) 
\cite{barman,mizokawani} are almost the same as that of 
La$_2$NiO$_4$ (Ni$^{2+}$) \cite{eisaki}.  

In order to confirm that the Ni $2p$ XAS and photoemission spectra are 
consistent with the Ni$^{2+} \rightarrow$ Ni$^{3+}$ 
valence change with La substitution, 
we have performed a configuration-interaction (CI) 
cluster-model calculation \cite{cluster1,Bocquet1}. 
Here, we considered a [NiO$_6$]$^{10-}$ octahedral cluster in the case of 
Ni$^{2+}$ and a [NiO$_6$]$^{9-}$ one in the case of Ni$^{3+}$. 
In this model, parameters to be fitted are the charge-transfer energy 
from the O $2p$ orbitals to the empty Ni $3d$ orbitals 
denoted by $\Delta$, the $3d-3d$ on-site Coulomb integration 
energy denoted by $U$, and the hybridization strength 
between the Ni $3d$ and O $2p$ orbitals denoted by 
Slater-Koster parameters denoted by $(pd\sigma)$ and 
$(pd\pi)$ \cite{JC}. 
The ratio of $(pd\sigma)/(pd\pi)$ is fixed to be 
$-2.2$ \cite{WA,STO1}. 
The configuration dependence of the transfer integrals 
has been taken into account \cite{DMS1}. 
Racah parameters B and C are fixed at 0.142 and 
0.527 eV, 80\% of the atomic Hartree Fock values \cite{Okada,deGroot}. 
We took into account the intra-atomic multiplet coupling 
for the Ni $2p$ XAS spectrum and the valence band spectrum, 
whereas the multiplet coupling between the core hole and 
the valence $d$ electrons was not taken into account for 
the Ni $2p$ photoemission spectrum as in the case of 
Ref.~\cite{Bocquet1}. 
The Slater integrals between the Ni $2p$ and $3d$ orbitals are 
$F^2=6.68$, $G^1=5.07$, and $G^3=2.88$ eV. 
The multiplet-averaged $2p$-$3d$ Coulomb interaction 
$Q(\equiv F^0-\frac{1}{15}G^1-\frac{3}{70}G^3)$ is 
$\sim$ 9.3 eV in the case of Ni$^{2+}$ and $\sim$ 
10 eV in the case of Ni$^{3+}$, satisfying the relationship 
$U/Q\sim 0.7$. 

The calculated Ni $2p$ XAS spectra have been 
broadened by a Gaussian and a Lorentzian. 
The full width at half maximum (FWHM) of the Gaussian is 
$\sim 1.0$ eV, which is determined by the instrumental 
resolution and the effect of 
electron-phonon coupling \cite{eph} 
and that of the Lorentzian is $\sim 0.5$ 
eV, which is determined from the natural width \cite{width}. 
The calculated Ni $2p$ core-level photoemission spectra 
have been broadened with an energy-dependent Lorentzian 
with a FWHM $2\Gamma = 2\Gamma_0(1+\alpha \Delta E)$, 
where $\Delta E$ denotes the energy separation from the main peak.  
We adopted the values $\alpha=0.15$ and $\Gamma_0=1.2$ eV. 
We then used a Gaussian broadening of 1.0 eV to simulate the 
instrumental resolution and broadening due to  
the core hole-$3d$ multiplet coupling. 

Figure \ref{NiXPS} shows that the Ni $2p$ 
photoemission spectra thus 
calculated reproduce the experimental Ni $2p$ photoemission 
spectra very well 
both in the cases of Ni$^{2+}$ and Ni$^{3+}$. 
These calculated spectra have been obtained 
with $\Delta = 4.5$ eV, $U=6.5$ eV, 
and $(pd\sigma)=-1.5$ eV in the case of Ni$^{2+}$ and 
$\Delta = 1.0$ eV, $U=7.0$ eV, 
and $(pd\sigma)=-1.1$ eV in the case of Ni$^{3+}$. 
These values are close to those previously reported : 
$\Delta = 4.5$ eV, $U=7.0$ eV, 
and $(pd\sigma)=-1.2$ eV in the case of La$_2$NiO$_4$ (Ni$^{2+}$) 
\cite{eisaki} and 
$\Delta = 1.0$ eV, $U=7.0$ eV, 
and $(pd\sigma)=-1.5$ eV in the case of PrNiO$_3$ (Ni$^{3+}$ or Ni$^{2+}\underline{L}$) 
\cite{mizokawani}. On the other hand, 
Fig.~\ref{Ni} shows that the calculated 
Ni $2p$ XAS spectrum for Ni$^{2+}$ is very different 
from that for Ni$^{3+}$. It should be noted that we have used the 
same parameter sets as in the case of Ni $2p$ photoemission spectra. 
The result thus 
explains why the Ni $2p$ XAS spectra change 
with the Ni valence, whereas the Ni $2p$ photoemission spectra do not. 

\begin{figure}
\begin{center}
\includegraphics[width=9cm]{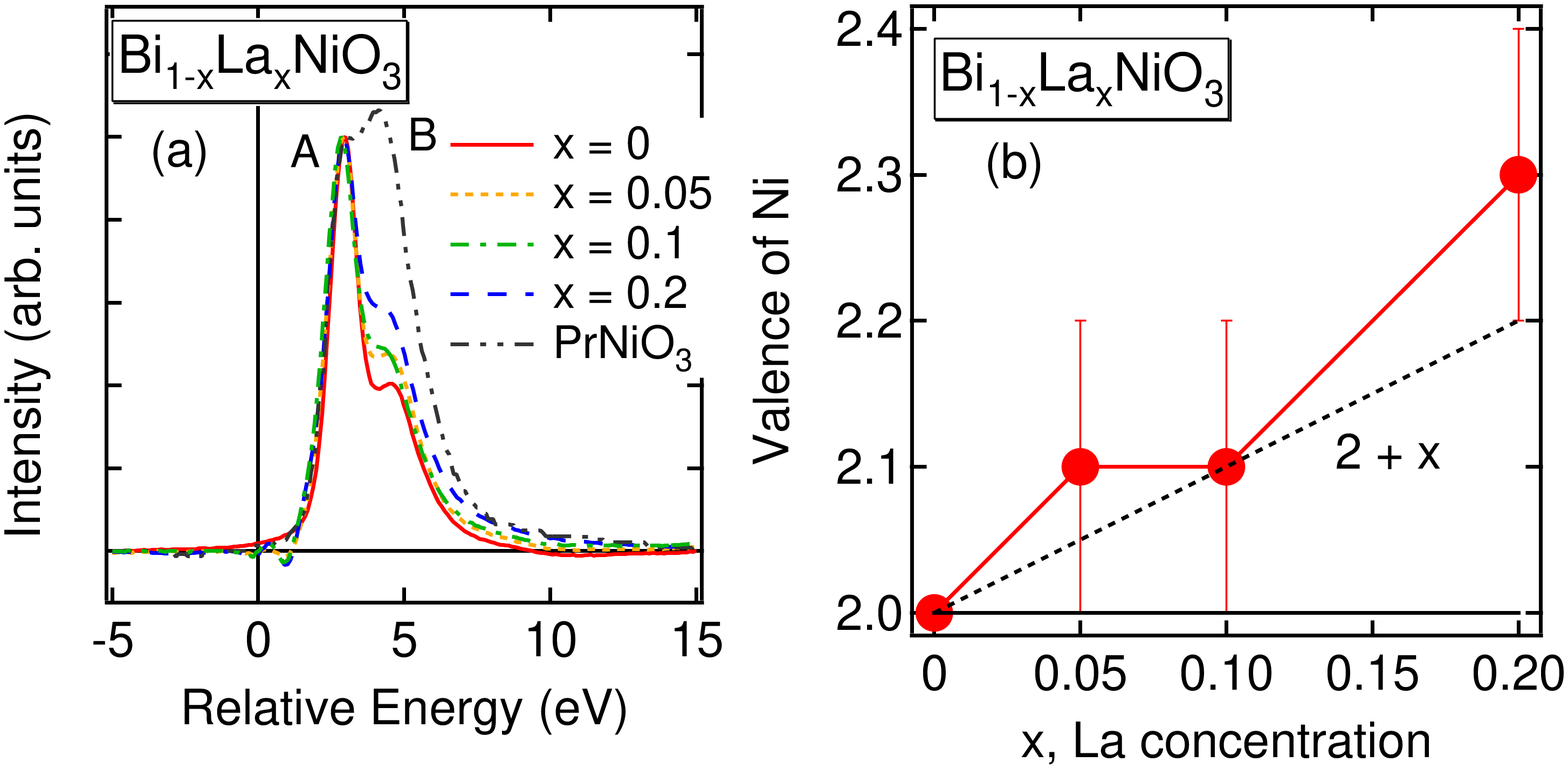}
\caption{(Color online) Ni $2p$ XAS spectra and the estimation of the valence of Ni. 
(a) Ni $2p$ XAS spectra after  the subtraction of background and the La $3d_{3/2}$ 
contribution. 
(b) Valence of Ni in Bi$_{1-x}$La$_x$NiO$_3$.} 
\label{Ni2}
\end{center}
\end{figure}

In order to evaluate the Ni valence more quantitatively 
from Ni $2p$ XAS, we have subtracted 
the contribution of La $3d_{3/2}$ XAS and 
compared the spectra with those of BiNiO$_3$ (Ni$^{2+}$) and 
PrNiO$_3$ (Ni$^{3+}$). 
Figure \ref{Ni2} (a) shows the Ni $2p$ XAS spectra 
after the subtraction of the background and the La $3d_{3/2}$ 
contribution. 
From the relative intensities 
of the two peaks, one can estimate the Ni 
valence as a function of $x$ as shown in Fig.~\ref{Ni2}. 
Here, we have assumed that the spectrum of $x=0$ represents pure Ni$^{2+}$, 
and adopted that of PrNiO$_3$ as a reference of Ni$^{3+}$. 
Figure 4 shows that 
the valence of Ni behaves approximately as 
$(2+x)+$, rather than $3+$, 
the value expected when the average valence 
of Bi becomes 3+ upon La substitution. 
The result indicates that La substitution does not change 
the average valence of Bi, 
while suppressing the charge disproportionation 
at the Bi site, and the valence of Ni becomes $\sim (2+x)+$. 
The valence of Bi shall be discussed below.

\subsection{O $1s$ and Bi $4f$ core-level photoemission spectra}
\begin{figure}
\begin{center}
\includegraphics[width=8cm]{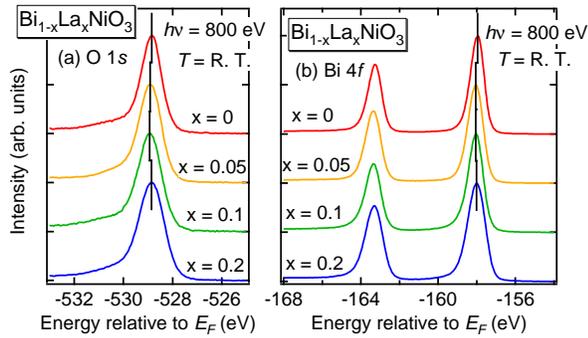}
\caption{(Color online) Core-level photoemission 
spectra of Bi$_{1-x}$La$_x$NiO$_3$. (a) O $1s$ (b) Bi $4f$.} 
\label{core}
\end{center}
\end{figure}

\begin{figure}
\begin{center}
\includegraphics[width=9cm]{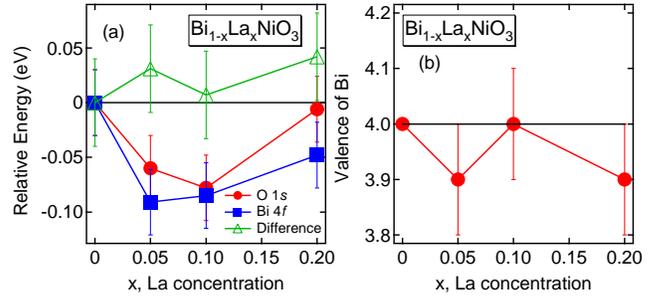}
\caption{(Color online) Estimation of the valence of Bi in 
Bi$_{1-x}$La$_x$NiO$_3$. 
(a) Binding energy shifts of the O $1s$ and Bi $4f$ core levels and 
their differences. 
(b) Valence of Bi calculated from the Ni valence deduced above.} 
\label{cor}
\end{center}
\end{figure}

Figure \ref{core} shows the core-level photoemission 
spectra of O $1s$ and Bi $4f$. 
Each O $1s$ spectrum consists of a single peak 
with negligible amount of contamination signals on the 
higher-binding-energy side, 
meaning that the surfaces of the samples 
became clean by scraping. 
The Bi $4f$ spectrum of BiNiO$_3$ does not show two components 
corresponding to Bi$^{3+}$ and Bi$^{5+}$, 
suggesting that the spectrum reflects 
the average valence of the Bi atom. 
The absence of a splitting into Bi$^{3+}$ and Bi$^{5+}$ components 
has been reported for BaBiO$_3$ \cite{BBO}, 
too, which was also
thought 
to exhibit a charge disproportionation 
into Bi$^{3+}$ and Bi$^{5+}$. 
The peak positions of both core-level spectra change with $x$. 
In Fig.~\ref{cor} (a), we have plotted the positions of 
the O $1s$ and Bi $4f$ core levels and 
their differences as functions of $x$.  
Both core levels are shifted downward up to $x\alt 0.1$ 
and then shifted upward from $x\sim 0.1$ to $x=0.2$.

The shift $\Delta E_B$ of a core-level binding energy with 
chemical composition measured 
from the chemical potential $\mu$ is given by 
\begin{equation}
 \Delta E_B=\Delta\mu+K\Delta Q+\Delta V_M+\Delta E_R,
\label{mm}
\end{equation}
where 
$\Delta\mu$ is the change in the chemical potential, 
$\Delta Q$ is the change in the number of valence electrons on the atom
considered, 
$\Delta V_M$ is the change in the Madelung potential, and 
$\Delta E_R$ is the change in the core-hole screening
\cite{Hufner-book}. 
The similar shifts of the O $1s$ and Bi $4f$ core levels indicate 
that the change in the Madelung potential ($\Delta V_M$) is negligibly
small because it would shift the core levels of anions and cations 
differently. 
Core-hole screening by conduction electrons is also considered to 
be negligibly small in transition-metal oxides \cite{Ino,Matsuno2,pote}.
Therefore, the result indicates that the chemical potential moves downward 
with increasing $x$, that is, hole doping takes place with La
substitution from $x\sim 0.1$ to $x=0.2$. 
For $x\alt 0.1$, however, 
the core levels are shifted in the opposite direction, 
probably reflecting the change of the band
structure caused by the considerable change in the lattice constants 
between $x=0$ and 0.05 \cite{BLNO}. 
The O $1s$ $-$ Bi $4f$ difference is 
free from the chemical potential shift and is 
considered to reflect the valence change of Bi called ``chemical
shift'' denoted by $K\Delta Q$ in Eq.~(\ref{mm}). 
From Fig.~\ref{cor} (a), 
the energy difference remains almost the same for all
values of $x$, indicating 
that the valence of Bi does not change appreciably. 
Figure \ref{cor} (b) shows that the valence of Bi calculated 
from the valence of Ni in Fig.~\ref{Ni2} (b) and 
the valence of La ($3+$) also does not 
show obvious $x$ dependence. 
Therefore, we consider that the valence of Bi remains 
almost unchanged, $4+$. 
Combining these results 
with the result of Ni $2p$ XAS, we conclude that 
the valence of Ni increases with increasing $x$, while 
the average valence of Bi remains constant, 
meaning 
that La substitution donates holes to the Ni sites. 
Charge transport in this system is therefore 
attributed to 
holes at both the Ni site (Ni$^{3+}$ or Ni$^{2+}\underline{L}$) 
and the Bi site (Bi$^{4+}$ or Bi$^{3+}\underline{L}$).
\subsection{Valence-band photoemission spectra}
\begin{figure}
\begin{center}
\includegraphics[width=9cm]{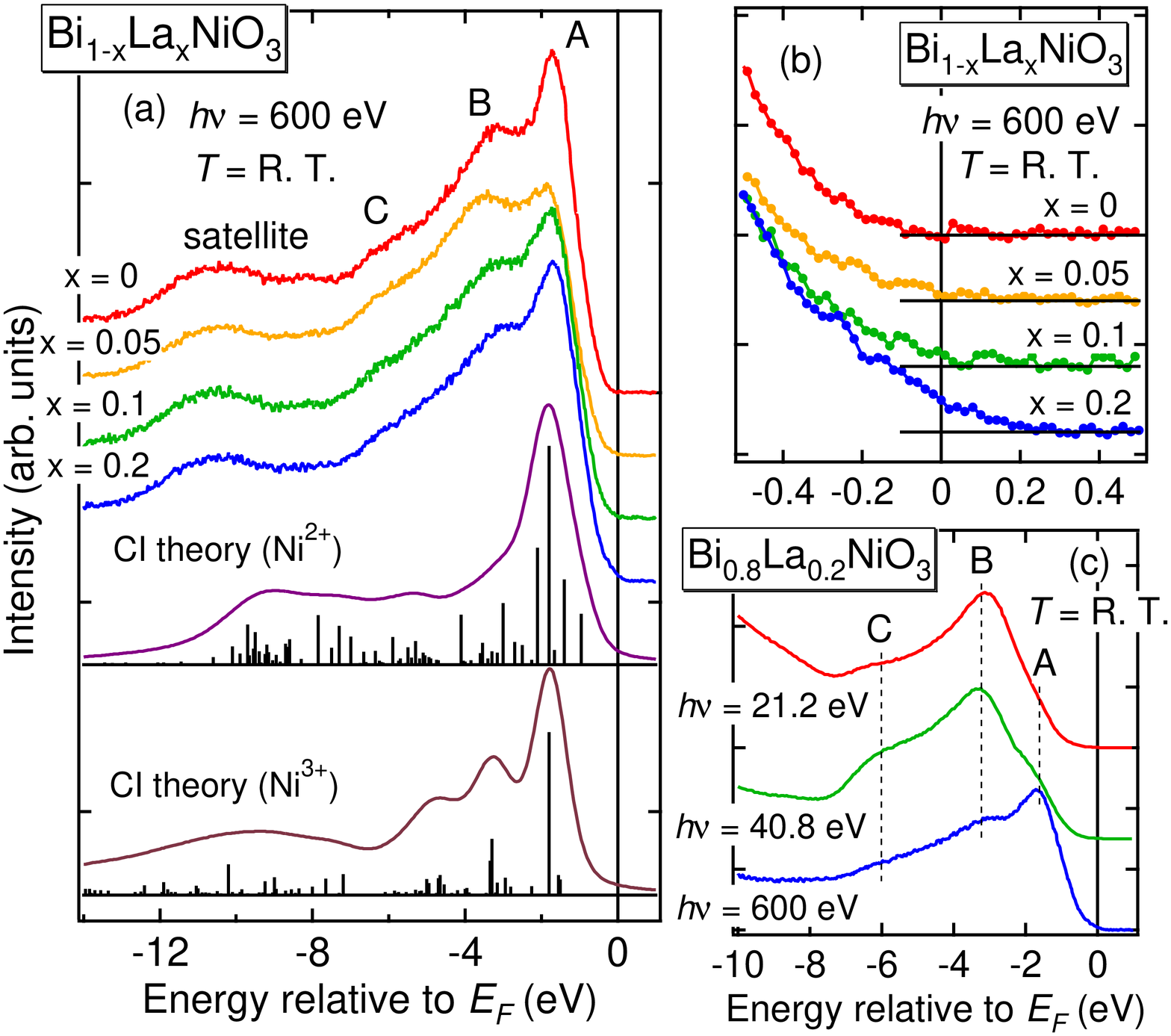}
\caption{(Color online) Valence-band spectra of Bi$_{1-x}$La$_x$NiO$_3$. 
(a) $h\nu = 600$ eV. The calculated spectra 
assuming Ni$^{2+}$ and Ni$^{3+}$ are presented at the bottom. 
(b) $h\nu = 600$ eV (near $E_F$). 
(c) $x=0.2$ taken at 21.2 eV, 40.8 eV, and 600 eV.} 
\label{val}
\end{center}
\end{figure}

The valence-band spectra of 
Bi$_{1-x}$La$_x$NiO$_3$ 
taken at 600 eV are shown in Fig.~\ref{val} (a). 
Considering the photoionization cross-sections \cite{Table}, 
most of the contributions come from Ni $3d$ 
and contributions 
from O $2p$ and Bi $6s$ are very small. 
(The relative photoionization cross-sections of 
the Ni $3d$, O $2p$, and Bi $6s$ orbitals are $\sim 19:1:1$ \cite{Table}.) 
One can observe three main structures labeled as A ($\sim -1.7$ eV), 
B ($\sim -3.3$ eV), and 
C ($\sim -6$ eV) and the satellite structure at $\sim -11$ eV. 
Figure \ref{val} (c) shows the photon energy dependence of 
the valence-band spectrum of $x=0.2$, including the spectra 
taken at $h\nu = 21.2$ eV and 40.8 eV. 
According to the photon-energy dependence of the photoionization 
cross-sections, 
the He I spectrum is dominated by the O $2p$ contribution 
(Ni $3d$ : O $2p$ : Bi $6s$ $\sim$ $0.37:1:0.003$.) 
and 
the Ni $3d$ contribution increases with increasing photon energy. 
Structure A, which is most pronounced at 600 eV, 
is therefore attributed to the contributions 
for Ni $3d$. 
On the other hand, structures B and C 
are due to the O $2p$ band. 
This assignment 
is in agreement with other Ni oxides like NiO \cite{cluster1} 
and $R$NiO$_3$ \cite{mizokawani}. 
We also performed CI cluster-model calculation by using the same
parameters as in the case of Ni $2p$ XAS and photoemission. 
At the bottom of Fig.~\ref{val} (a) are shown the calculated spectra for 
the [NiO$_6$]$^{10-}$ cluster (Ni$^{2+}$) and 
the [NiO$_6$]$^{9-}$ cluster (Ni$^{3+}$). 
The calculated valence-band 
spectra have been broadened with a Gaussian of 0.8 eV FWHM 
to account for the combined effects of the instrumental 
resolution and the $d$ band dispersion, and an 
energy-dependent Lorentzian ($\mbox{FWHM}=0.2|E-E_F|$) 
(Ref.~\cite{Hufner}) to account for the lifetime broadening 
of the photohole. 
Considering the photoionization cross-sections, the Ni $3d$ emission 
overwhelms the O $2p$ emission at $h\nu = 600$ eV, where 
three main structures and the satellite structure are well 
reproduced in the 
calculation for both Ni$^{2+}$ and Ni$^{3+}$. 

Spectra in the vicinity of $E_F$ are shown in Fig.~\ref{val} 
(b). There is no $E_F$ cutoff for the $x=0$ 
sample, consistent with the insulating behavior. 
Upon La substitution, emission appears at $E_F$ 
and increases with increasing $x$, indicating the 
insulator-to-metal transition induced by  
La substitution. 
\section{Conclusion}
In conclusion, 
we have measured the photoemission and x-ray 
absorption spectra of Bi$_{1-x}$La$_{x}$NiO$_{3}$. 
From Ni $2p$ XAS, we observed the change of the valence of
Ni. Combined with core-level photoemission studies, 
it was found that 
the average valence of Bi remains $\sim 4+$ with increasing $x$, 
and La substitution causes hole doping into the Ni sites. 
These results mean that both the Bi 
and Ni sites contribute to the metallic conductivity. 
From the valence-band spectra, 
we observed a Fermi-level cutoff for $x>0$.  
The spectra of the Ni $2p$ XAS, the Ni $2p$ photoemission, and the valence band  
have been analyzed by configuration-interaction cluster-model
 calculation, and good agreement between experiment and calculation 
was obtained. 
\section*{Acknowledgment}
The authors would like to thank K. Ono and A. Yagishita 
for their support at KEK-PF. 
This work was supported by a Grant-in-Aid for Scientific Research 
(S17105002) 
from the Ministry of Education, Culture, Sports, 
Science and Technology, Japan.
This work was done under the approval of the Photon Factory 
Program Advisory Committee (Proposal No. 2003G149). 
H. W. acknowledges support from 
the Japan Society for the Promotion of Science for Young 
Scientists.
\bibliography{BLNOtex}
\end{document}